\documentclass[12pt]{article}             % Use with LaTeX2e

\usepackage{amsmath,amssymb,amsthm}
\usepackage{mathbbol,bbm}

\swapnumbers
\theoremstyle{definition}
\newtheorem{defin}{Definition}[section]

\renewcommand{\H}{\mathcal{H}}
\newcommand{\Hin}{\H_{\mathrm{in}}}
\newcommand{\Hout}{\H_{\mathrm{out}}}
\newcommand{\T}{\mathcal{T}}
\renewcommand{\S}{\mathcal{S}}
\newcommand{\map}{\Phi}
\newcommand{\mapp}{\Lambda}
\newcommand{\cm}{C(\map)}
\newcommand{\la}{\langle}
\newcommand{\ra}{\rangle}
\newcommand{\B}{{\mathcal B}}
\newcommand{\id}{\r{id}}
\newcommand{\Cx}{\mathbbm{C}}
\newcommand{\Ad}{\mathrm{Ad}}
\newcommand{\bi}[1]{\boldsymbol{#1}}
\newcommand{\idty}{\Eins}
\DeclareMathOperator{\Tr}{Tr}
\providecommand{\abs}[1]{\lvert#1\rvert}
\providecommand{\norm}[1]{\lVert#1\rVert}
\newcommand{\s}[1]{\mathsf{#1}}
\renewcommand{\r}[1]{\mathrm{#1}}

\begin{document}

\begin{center}
\textbf{\large
 Additivity of minimal entropy output for a class of covariant channels} \\[6pt]
{\large
 M.~Fannes\footnote{E-mail: mark.fannes@fys.kuleuven.ac.be},
 B.~Haegeman\footnote{E-mail: bart.haegeman@fys.kuleuven.ac.be, Research
assistant of the Fund for Scientific Research--Flanders (Belgium)
(F.W.O.--Vlaanderen)},
 M.~Mosonyi\footnote{E-mail: mosonyi@math.bme.hu,
 On leave of absence from Mathematical Institute, Budapest University of Technology and
 Economics, 1111 Budapest XI. Egry Jozsef u. 1., Hungary}}
 and
{\large
 D.~Vanpeteghem\footnote{E-mail: dimitri.vanpeteghem@fys.kuleuven.ac.be, Research
assistant of the Fund for Scientific Research--Flanders (Belgium)
(F.W.O.--Vlaanderen)}} \\[6pt]
\emph{Instituut voor Theoretische Fysica, K.U.Leuven, Celestijnenlaan~200D,
B-3001 Heverlee, Belgium}
\end{center}
\bigskip

\noindent
\textbf{Abstract:}
Additivity of minimal entropy output is proven for the class of quantum channels
$\mapp_t (A):=t A^{\s T}+(1-t)\tau(A)$ \
in the parameter range \ $-2/(\r d^2-2)\le t\le 1/(\r d+1)$.
\medskip

\section{Introduction}

A quantum channel $\Phi$ is a CPTP (completely positive, trace-preserving)
map that maps states of the input Hilbert space $\Hin$ into states of the
output Hilbert space $\Hout$. The minimal entropy output of the channel is
\begin{equation*}
 \s S_{\mathrm{min}}(\map) := \min\bigl\{\s S\bigl(\map(\rho)\bigr) \,:\,
 \rho\in\S(\Hin)\bigr\},
\end{equation*}
where $\S(\Hin)$ denotes the state space of the input system, i.e.\ the set
of self-adjoint matrices with trace one, and $\s S(\sigma) := -\Tr
\sigma\log\sigma$ is the von Neumann entropy of the density matrix $\sigma$.
By concavity of the entropy, $\s S_{\mathrm{min}}$ can be expressed as
\begin{equation*}
 \s S_{\mathrm{min}}(\map) := \min\bigl\{ \s S\bigl(
 \map(|\psi\ra\la\psi|)\bigr) \,:\, \psi\in\Hin,\ \norm{\psi}=1 \bigr\}.
\end{equation*}
Hence, $\s S_{\mathrm{min}}$ is a lower bound on the mixedness of the output
state, measured by entropy, when the input is a pure state. It can therefore
be interpreted as a measure of the noisiness of the channel.

A fundamental question of quantum information theory is whether the minimal
entropy output is additive for product channels, i.e., given two channels
$\map_1: \S(\Hin^1) \to \S(\Hout^1)$ and
$\map_2: \S(\Hin^2) \to \S(\Hout^2)$, whether the equality
\begin{equation}
 \s S_{\mathrm{min}}(\map_1\otimes\map_2) = \s S_{\mathrm{min}}(\map_1) +
 \s S_{\mathrm{min}}(\map_2)
\label{addit}
\end{equation}
holds. This question was shown to be equivalent to the additivity of other
quantities, like the Holevo capacity and the entanglement of
formation~\cite{MSW,AB,Shor1,Shor2}. It has been solved in the special cases
when one of the channels is the identical channel~\cite{AHW}, the
depolarising channel~\cite{King3}, a unital qubit channel~\cite{King2} or an
entanglement breaking channel~\cite{Shor3,King}.

The inequality
\begin{equation*}
 \s S_{\mathrm{min}}(\map_1\otimes\map_2) \le \s S_{\mathrm{min}}(\map_1) +
 \s S_{\mathrm{min}}(\map_2)
\end{equation*}
holds trivially, since $\s S\bigl( \map_1\otimes\map_2(|\psi_1\otimes\psi_2\ra
\la\psi_1\otimes\psi_2|) \bigr) = \s S_{\mathrm{min}}(\map_1) +
\s S_{\mathrm{min}}(\map_2)$ for the states $|\psi_1\ra \la\psi_1|$ and
$|\psi_2\ra \la\psi_2|$ at which the output entropies of $\map_1$ and $\map_2$
are minimised. Therefore, in order to show additivity, we have to show that
$\s S\bigl( \map_1\otimes\map_2(|\psi\ra \la\psi|) \bigr)$ attains its
minimal value at $\psi = \psi_1\otimes\psi_2$. Using the Schmidt decomposition
of $\psi$, we have to minimise
\begin{equation}
 \s S\biggl( \sum_{ij} \sqrt{\lambda_i\lambda_j}\, \map_1(|e_i\ra\la e_j|)
 \otimes \map_2(|f_i\ra\la f_j|) \biggr),
\label{entropy}
\end{equation}
where $\{e_i \,:\, i=1,2,\ldots, \dim\Hin^1\}$ and $\{f_j \,:\, j=1,2,\ldots,
\dim\Hin^2\}$ are orthonormal bases of the input spaces $\Hin^1$ and $\Hin^2$
and where $\bi\lambda = \{\lambda_i \,:\, i=1,\ldots, \bigl(\dim\Hin^1\bigr) 
\wedge \bigl(\dim\Hin^2\bigr)\}$ is a probability distribution.

Solving this problem in full generality is rather hopeless, however, it
simplifies a great deal if the channels $\map_1$ and $\map_2$
possess the additional property of unitary covariance.

\begin{defin}
 A channel $\map: \S(\Hin)\to\S(\Hout) $ is unitarily covariant, if for every
 unitary $U$ in $\Hin$ there exists a unitary $V$ in $\Hout$ such that
 \begin{equation}
   \map\circ\Ad_U = \Ad_V\circ\map
 \label{covariance}
 \end{equation}
 holds, where $\Ad_X$ is the map $A \mapsto XAX^*$.
\end{defin}

Obviously, for unitarily covariant channels $\s S\bigl( \map(|\psi\ra
\la\psi|) \bigr)$ is independent of the input state $|\psi\ra \la\psi|$
and therefore equal to $\s S_{\mathrm{min}}(\map)$.

A significant simplification of the additivity problem for unitarily
covariant maps is that the entropy in~(\ref{entropy}) depends only on the
Schmidt coefficients of the input state $\psi$. Indeed, if $\psi$ and
$\varphi$ have the same Schmidt coefficients, then there exist unitaries
$U_1$ and $U_2$ in $\Hin^1$ and $\Hin^2$ such that $\varphi = U_1\otimes U_2\,
\psi$. Therefore
\begin{equation*}
 \map_1\otimes\map_2( |\varphi\ra \la\varphi|) = (V_1\otimes V_2)\,
 \map_1\otimes\map_2(|\psi\ra \la\psi|)  (V_1^*\otimes V_2^*)
\end{equation*}
for some unitaries $V_1$ and $V_2$ in $\Hout^1$ and $\Hout^2$, hence
$\map_1\otimes\map_2( |\varphi\ra \la\varphi|)$ and
$\map_1\otimes\map_2( |\psi\ra \la\psi|)$ have the same entropy. Now
to prove additivity, it is sufficient to show that the function
\begin{equation*}
 \bi\lambda \mapsto \s S\biggl( \sum_{ij} \sqrt{\lambda_i\lambda_j}\,
 \map_1(|e_i\ra\la e_j|) \otimes \map_2(|f_i\ra\la f_j|) \biggr)
\end{equation*}
reaches its minimal value at the vertices of the simplex $\S$. Here, $\{e_i\}$
and $\{f_j\}$ are fixed orthonormal bases in $\Hin^1$ and $\Hin^2$ and
$\S$ is the simplex of probability distributions on $\bigl(\dim\Hin^1\bigr) 
\wedge \bigl(\dim\Hin^2\bigr)$ points.

A trivial example of a unitarily covariant channel is the trace map
\begin{equation*}
 \tau(A) := (\Tr A)\, \frac{1}{\r d_{\r{out}}}\, \idty_{\r{out}},
\end{equation*}
where $\idty_{\r{out}}$ is the identity on $\Hout$ and
$\r d_{\r{out}} := \dim\Hout$. In this case, the covariance relation~(\ref{covariance})
holds for any pair of unitaries $(U,V)$. However, if we strengthen the definition
of covariance, and require uniqueness of $V$, then the map
$U\mapsto V$ is easily seen to be a representation of the group $\mathcal G =
\s U(\r d_{\r{in}})$ on $\Hout$. As $\Ad_U$ is insensitive to phase
factors, we can consider $\mathcal G = \s{SU}(\r d_{\r{in}})$ as well. We can
now turn the question around, fixing a unitary representation $\alpha_{\r{out}}$ of
$\s{SU}(\r d_{\r{in}})$ on $\Hout$ and looking for channels that satisfy the
following property:

\begin{defin}\label{cov2}
 A channel $\map: \S(\Hin)\to\S(\Hout)$ is $\alpha$-invariant, if
 \begin{equation*}
   \map\circ\Ad_U = \Ad_{\alpha_{\r{out}}(U)}\circ\map
 \end{equation*}
 for all $U \in \s{SU}(\r d_{\r{in}})$.
\end{defin}

Such $\alpha$-invariant channels were studied in~\cite{Holevo}, where it was also
shown that for an irreducible representation $\alpha_{\r{out}}$
\begin{equation*}
 \chi(\map) = \log \r d_{\r{out}} - \s S_{\mathrm{min}}(\map).
\end{equation*}
Here, $\chi(\map)$ is the Holevo capacity of the channel~\cite{Holevo2}, given by the formula
\begin{equation*}
 \chi(\map) = \max \Bigl\{ \s S\Bigl( \sum_j p_j \map(\rho_j)\Bigr) - \sum_j p_j
 \s S\bigl( \map(\rho_j) \bigr) \Bigr\},
\end{equation*}
where the maximum is taken over all finitely supported measures $\bf p$ on the state
space $\S(\Hin)$ and ${\bf p}(\rho_j) =: p_j$.
Therefore, for such channels, if additivity holds for the minimal entropy output it also holds
for the Holevo capacity.

In section~\ref{inv}, we deal with the description of the set $\B_{\alpha}$ of
$\alpha$-invariant channels, focusing mainly on the case $\Hout=\Hin$ and
$\alpha_{\r{out}}$ is
either
\begin{itemize}
\vspace*{-12pt}
\item[(a)]
 the identical representation, or
\item[(b)]
\vspace*{-6pt}
the conjugate representation $U\mapsto \overline U$.
\end{itemize}
\vspace*{-9pt}
$\overline U$ is the entrywise conjugate of the matrix $U$ in some fixed
base. It turns out that
\begin{itemize}
\vspace*{-12pt}
\item[$\bullet$]
 in case~(a),
 $\B_{\alpha}$ is the family of depolarising channels, given by
 \begin{equation*}
 \Delta_t (A) := tA + (1-t)\tau(A),\qquad  -\frac{1}{\r d^2-1}\le t\le 1,
 \end{equation*}
 where $\r d = \r d_{\r{in}} = \r d_{\r{out}}$ and $[-1/(\r d^2-1), 1]$ is the
 maximal range of $t$ such that $\Delta_t$ is completely positive, while
\item[$\bullet$]
\vspace*{-6pt}
 in case~(b),
 we obtain the family
 \begin{equation*}
 \mapp_t(A) := t A^{\s T} + (1-t)\tau(A),\qquad  -\frac{1}{\r d-1}\le t\le
\frac{1}{\r d+1},
\end{equation*}
 where $A^{\s T}$ is the transpose of $A$ in the same base in which the $U$'s are
 conjugated.
\end{itemize}
\vspace*{-12pt}
The elements of this second class will be called transpose depolarising channels.

Additivity of the minimal entropy output was shown for the depolarising channels
in~\cite{King3}, even when one of the channels is completely arbitrary, and for
the extreme transpose depolarising channel $\mapp_{-1/(\r d-1)}$ in~\cite{MY},
followed by proofs using different methods in~\cite{DHS} and~\cite{AF}. In
section~\ref{meo}, we shall show additivity for the subset of the transpose
depolarising channels determined by $-2/(\r d^2-2)\le t \le 1/(\r d+1)$.

\section{Extreme invariant maps}
\label{inv}

We start out with the more general setting of a pair $\alpha$ of unitary
representations $\alpha_{\r{in}}$ and $\alpha_{\r{out}}$
of an abstract group $\mathcal G$ on $\Hin$ and $\Hout$ respectively and
define a map $\map$ to be $\alpha$-invariant, if
\begin{equation*}
 \map\circ\Ad_{\alpha_{\r{in}}(g)} =
 \Ad_{\alpha_{\r{out}}(g)}\circ\map,\qquad g\in G.
\end{equation*}

Next, we fix a basis $\{e_i \,:\, i=1,2,\ldots ,d_{\r{in}}\}$ in $\Hin$
with corresponding anti-unitary $J$ defined by $Je_i=e_i$. The map
\begin{equation*}
 \T: |B\ra\la A| \mapsto JAJ^*\otimes B
\end{equation*}
is well-defined on $\B\bigl(\B(\Hin), \B(\Hout)\Bigr)$ and extends to a
linear isomorphism between $\B\bigl(\B(\Hin),\B(\Hout)\bigr)$ and
$\B(\Hin) \otimes \B(\Hout)$. For a general element $\map:
\B(\Hin)\to\B(\Hout)$ we obtain the formula
\begin{equation}
 \T\map = \sum_{ij} e_{ij}\otimes\map(e_{ij}).
\label{choimatrix}
\end{equation}
The right-hand side of eq.~(\ref{choimatrix}) is called the Choi matrix
of $\map$ and we shall denote it by $C(\map)$. A fundamental theorem of
Choi~\cite{C} states that $\map$ is completely positive if and only if
its Choi matrix is positive semidefinite.

A straightforward computation shows that $\map$ is trace-preserving if
and only if $\Tr_2\,\cm = \idty_{\r{in}}$ and, moreover,
\begin{align*}
 &(\text{i})\quad C(\Ad_V\circ\map) = \Ad_{\idty\otimes V}\bigl( C(\map) \bigr)
 = (\idty\otimes V)\ C(\map)\ (\idty\otimes V^*) \\
 &(\text{ii})\quad C(\map\circ\Ad_U) = \Ad_{JU^*J^* \otimes \idty}\bigl( C(\map)
 \bigr) \\
 &\phantom{(\text{ii}\quad) C(\map\circ\Ad_U)} = (JU^*J^* \otimes \idty)\
 C(\map)\ (JUJ^* \otimes \idty).
\end{align*}
Properties~(i) and~(ii) imply that $\map$ is $\alpha$-invariant if and only if
\begin{equation*}
 \bigl[ C(\map) \,,\, J\alpha_{\r{in}}(g)J^* \otimes \alpha_{\r{out}}(g)
 \bigr] = 0,\qquad g\in\mathcal G.
\end{equation*}

The map $g \mapsto \kappa(g) := J\alpha_{\r{in}}(g)J^* \otimes
\alpha_{\r{out}}(g)$ is
again a unitary representation of $\mathcal G$, which decomposes in
irreducible components
\begin{equation*}
 \kappa(g) \cong \bigoplus_k \kappa_k(g) \otimes \idty_{\r m(k)},
\end{equation*}
where $k$ labels inequivalent irreducible representations of $\mathcal G$ on a
$\r d(k)$ dimensional Hilbert space and where $\r m(k)$ is the multiplicity of
the $k$-th representation. A general element of the commutant of
$\kappa\bigl( \s{SU}(\r d_{\r{in}})\bigr)$ is of the form
\begin{equation*}
 C = \bigoplus_k \idty_{\r d(k)} \otimes A_k,
\end{equation*}
and it is the Choi matrix of a CP map if and only if all $A_k\ge 0$.
In the special case $\r m(k)=1\ \forall k$, we obtain that the map $\map$ is
$\alpha$-invariant if and only if its Choi matrix has the form
\begin{equation}
 \cm = \sum_k c_k\,P_k,
\label{decomp}
\end{equation}
where $P_k$ is the projection onto the Hilbert space of the $k$-th irreducible
component and where the $c_k$'s are arbitrary complex numbers. Obviously,
$\map$ is CP if and only if all the coefficients $c_k$ are nonnegative.
The maps $\T^*P_k$ span the extreme rays of the convex cone of $\alpha$-invariant
CP maps. However, $\T^* P_k$ need not be trace-preserving, therefore
formula~(\ref{decomp}) is in general insufficient to describe the convex set
$\B_{\alpha}$ of $\alpha$-invariant CPTP maps.

Note that the map
\begin{equation*}
 \frac{1}{\r d_{\r{in}}}\,\T: \map \mapsto \frac{1}{\r d_{\r{in}}}\,\cm
\end{equation*}
gives a proper embedding of the convex set of CPTP maps into the state
space of $\Hin\otimes \Hout$, its range, however, does not coincide with
the whole
of the state space because of the restriction $\Tr_2\,\cm =
\idty_{\r{in}}$. Thus $\alpha$-invariant CPTP maps can be identified
with a subset of the set $\S_\alpha$ of $\alpha$-invariant states,
i.e.\ states that satisfy $\bigl[\rho \,,\, J\alpha_{\r{in}}(g)J^* \otimes
\alpha_{\r{out}}(g) \bigr] = 0\ \forall g\in \mathcal G$. The
characterisation of $\S_\alpha$ for different symmetry groups was
studied in~\cite{VW}. Even if in general $\S_{\alpha}$ is strictly
larger then the embedded image of $\B_\alpha$, both sets coincide when
$\alpha_{\r{in}}$ is irreducible since the invariance
\begin{align*}
  \Tr_2 \rho
  &= \Tr_2 \bigl( J\alpha_{\r{in}}(g)J^* \otimes \alpha_{\r{out}}(g)
  \bigr)\, \rho\, \bigl( J\alpha_{\r{in}}(g)^*J^* \otimes
  \alpha_{\r{out}}(g)^* \bigr) \\
  &= \bigl( J\alpha_{\r{in}}(g)J^* \bigr)\, \bigl(\Tr_2 \rho\bigr)\,
  \bigl( J\alpha_{\r{in}}(g)^*J^* \bigr)
\end{align*}
implies $\Tr_2 \rho = \frac{1}{\r d_{\r{in}}} \idty_{\r{in}}$.

Now we turn to the case $\Hin=\Hout=\H$, $\alpha_{\r{in}}$ the identical
representation of $\mathcal G = \s{SU}(\r d)$ and either
$\alpha_{\r{out}}(U) = U$, case~(a), or $\alpha_{\r{out}}(U) =
\overline{U} = JUJ^*$, case~(b). \\
In case~(a), the invariant projections of $JUJ^*\otimes U$
are $P := |\psi\ra\la\psi|$ and $P^\perp$, where $\psi =
\frac{1}{\sqrt{\r d}} \sum_i e_i\otimes e_i$. The corresponding extreme
maps are
\begin{equation*}
 \T^*(\r dP) = \Delta_1 = \id
 \qquad\text{and}\qquad
 \T^*\Bigl( \frac{\r d}{\r d^2-1}\, P^\perp \Bigr) = \Delta_{-1/(\r d^2-1)}.
\end{equation*}
In case~(b), the invariant projections are $P_{\r s}$ and $P_{\r a}$, the
projections onto the symmetric and the antisymmetric subspaces of
$\H\otimes\H$. The corresponding extreme maps are
\begin{equation*}
 \T^*\Bigl( \frac{2}{\r d+1}\, P_{\r s} \Bigr) = \mapp_{1/(\r d+1)}
 \qquad\text{and}\qquad
 \T^*\Bigl( \frac{2}{\r d-1}\, P_{\r a} \Bigr) = \mapp_{-1/(\r d-1)}.
\end{equation*}
For details on decomposing tensor products of representations of
$\s{SU}(\r d)$, see e.g.\ \cite{HFJ}.

\section{Minimal entropy output for the transpose depolarising channel}
\label{meo}

In this section, we shall prove that
\begin{equation}
 \s S_{\r{min}}\bigl( \mapp_t\otimes\mapp_t \bigr) = 2\
 \s S_{\r{min}}\bigl( \mapp_t \bigr),\qquad -\frac{2}{\r d^2-2}\le t\le
 \frac{1}{\r d+1}.
\label{add2}
\end{equation}

First note that the partial transpose of the state $\frac{1}{\r d}
C\bigl(\mapp_t\bigr)$ is $\frac{1}{\r d} C\bigl(\Delta_t\bigr)$, hence
it is positive if and only if $t\in\bigl[-1/(\r d^2-1) \,,\, 1/(\r d+1)
\bigr] =: R_1$. Having positive partial transpose is a necessary condition
for a bipartite state to be separable~\cite{Peres} and was shown
in~\cite{VW} to be also sufficient for unitarily covariant states .
Quantum channels with separable Choi matrix are called entanglement-breaking,
and additivity of their minimal entropy output was proven
in~\cite{Shor3}. It holds even in the stronger form where one of the factors
in~(\ref{addit}) is entanglement-breaking and the other is completely
arbitrary. Hence, additivity of minimal entropy output holds in the strong form
\begin{equation}
 \s S_{\r{min}}\bigl( \mapp_t\otimes\map \bigr) = \s S_{\r{min}}\bigl(
 \mapp_t \bigr) + \s S_{\r{min}}\bigl( \map \bigr),
\label{add}
\end{equation}
when $t\in R_1$ and $\map$ is an arbitrary quantum channel. We can
therefore assume $t$ to be negative for the rest of the argument, even though
the proof applies with a trivial modification to positive $t$'s as well.

Let $\rho(\bi\lambda) := |\psi(\bi\lambda)\ra \la\psi(\bi\lambda)| = \sum_{ij}
\sqrt{\lambda_i\lambda_j}\, |e_i\ra\la e_j| \otimes |e_i\ra\la e_j|$ denote
the input state and $X(\bi\lambda) := \mapp_t\otimes\mapp_t\bigl(
\rho(\bi\lambda) \bigr)$ the output state for some fixed $t\le0$. An easy
computation shows that
\begin{equation*}
 X(\bi\lambda) = t^2\,\rho(\bi\lambda) + \frac{t(1-t)}{\r d}\bigl(
 \rho'(\bi\lambda)\otimes\idty + \idty\otimes\rho'(\bi\lambda) \bigr) +
 \Bigl( \frac{1-t}{\r d} \Bigr)^2\, \idty\otimes\idty,
\end{equation*}
where $\rho'(\bi\lambda) := \Tr_2 \rho(\bi\lambda) = \Tr_1 \rho(\bi\lambda)$.
The basis $\{e_i\}$ is the one in which the transpose is taken.

The vectors $\{e_i\otimes e_j \,:\, i\ne j\}$ are eigenvectors of $X(\bi\lambda)$
with corresponding eigenvalues $\eta_{ij} = t(1-t)(\lambda_i
+ \lambda_j)/\r d + \bigl((1-t)/\r d\bigr)^2$. The contribution of this part of the
spectrum to the entropy is $\s S_1(\bi\lambda) = -\sum_{i\ne j}
\eta_{ij}\ln\eta_{ij}$, which, being a concave function of $\bi\lambda$, takes
its minimal value at the vertices of $\S$.

Now we consider the subspace spanned by $\{ e_i\otimes e_i\,:\,
i=1,\ldots ,\r d\}$. The restricted output density has the form
\begin{equation*}
 \hat X(\bi\lambda) = t^2\, |\sqrt{\bi\lambda}\ra \la\sqrt{\bi\lambda}| +
 \frac{2t(1-t)}{\r d} \r{diag}(\bi\lambda) + \Bigl( \frac{1-t}{\r d}
 \Bigr)^2  \idty\otimes\idty,
\end{equation*}
where $|\sqrt{\bi\lambda}\ra = \sum_j \sqrt{\lambda_j}\,e_j$ and where
$\r{diag}(\bi\lambda)$ is the diagonal matrix with diagonal $\bi\lambda$.
Note that, at the vertices of $\S$, $\lambda_j = \delta_{jk}$ for some $k$
and that in this case $\hat X(\bi\lambda)$ is diagonal in the basis
$\{ e_i\otimes e_i\}$. Let $\s S_2(\bi\lambda)$ denote the entropy contribution
of the eigenvalues of $\hat X(\bi\lambda)$. In order to show that it takes
its minimal value at the vertices of $\S$ it is sufficient to show
that $\hat X(\bi\lambda)$ is more mixed than $\hat X(\bi\lambda^*) $ for any
$\bi\lambda\in\S$, where $\lambda^*_j = \delta_{j,1}$.

Recall that, for hermitian matrices, $A_1$ is called more mixed than
$A_2$, in notation $A_1\succ A_2$, if
\begin{align*}
 \kappa_1(A_1)
 &\le \kappa_1(A_2) \\
 \kappa_1(A_1) +  \kappa_2(A_1)
 &\le \kappa_1(A_2) + \kappa_2(A_2) \\
 &\ \vdots \\
 \Tr A_1
 &= \Tr\,A_2.
\end{align*}
Here, $\kappa_1(A_k) \ge\kappa_2(A_k) \ge\ldots$ are the eigenvalues of $A_k$
in decreasing order. Note that $\sum_{i\neq j}\eta_{ij} = (1-t^2)(\r
d-1)/2\r d$ for any $\bi\lambda$ and that therefore $\Tr \hat
X(\bi\lambda) = 1 - \sum_{i\neq j} \eta_{ij}$ is independent of $\bi\lambda$. It is
also clear that the $\bi\lambda$-independent diagonal term
$\bigl((1-t)/\r d\bigr)^2 \idty\otimes\idty$ doesn't influence the more
mixedness relation. The eigenvalue vector of
$\hat X(\bi\lambda^*) - \bigl((1-t)/\r d\bigr)^2 \idty\otimes\idty$, arranged in
decreasing order, is $\bigl( 0,0,\ldots, 0,t(2+(d-2)t)/\r d
\bigr)$. We have therefore to show that $\hat X(\bi\lambda) - \bigl(
(1-t)/\r d \bigr)^2 \idty\otimes\idty$ is negative semidefinite.
Writing out explicitly the corresponding quadratic form, this criterion
becomes
\begin{equation*}
 \biggl| \sum_j \sqrt{\lambda_j}\,x_j \biggr|^2 \le -\frac{2(1-t)}{t\r d} \sum_j
 \lambda_j\,\abs{x_j}^2,\qquad (x_1,\ldots ,x_{\r d})\in\Cx^{\r d},
\end{equation*}
which, by the Schwarz inequality, holds if and only if $\r d \le
- 2(1-t)/t\r d$, that is, $-2/(\r d^2-2)\le t$.

The case $\r d=2$ is special in the sense that the conjugate representation of
$\s{SU}(\r d)$ is unitarily equivalent to the identical representation.
As a consequence, the family of transpose depolarising channels coincides with
the family of depolarising channels and additivity of the minimal entropy
output holds in the stronger form~(\ref{add}) for all $t$ allowed by complete
positivity. Also, the irreducibility of $\alpha_{\r{out}}$ implies that an
$\alpha$-invariant channel is bistochastic when $\r d_{\r{in}} = \r
d_{\r{out}}$ and therefore~(\ref{add}) follows also from~\cite{King2}
when $\r d=2$. Yet another proof of the weak version~(\ref{add2}) of additivity
can be given in this case by showing that
\begin{equation}
 \bi\lambda'\succ\bi\lambda
 \qquad\text{implies}\qquad
 X(\bi\lambda')\succ X(\bi\lambda),
\label{mm}
\end{equation}
for any $\bi\lambda,\,\bi\lambda'\in\S$, which is a matter of straightforward
computation and immediately yields~(\ref{add2}). Numerical computations
suggest that~(\ref{mm}) holds for $\r d>2$ and arbitrary value of $t$ as well,
which would clearly be sufficient to prove~(\ref{add2}) for the missing
domain $t\in\bigl[ -1/(\r d-1) \,,\, -2/(\r d^2-2) \bigr]$.

\noindent
\textbf{Acknowledgements: }
This work was partially supported by F.W.O. Vlaanderen grant G.0109.01.

\end{document}